\documentclass[aps,prb,twocolumn,superscriptaddress]{revtex4}
\usepackage{bm}
\usepackage{amsmath}
\usepackage{amssymb}
\usepackage{graphicx}
\usepackage{color}
\input{epsf}
\begin{document}
\title{Spin Gating of Mesoscopic Devices}
\author{R. I. Shekhter\footnote{Corresponding author. E-mail: shekhter@physics.gu.se}}
\affiliation{Department of Physics, University of Gothenburg,
SE-412 96 G{\" o}teborg, Sweden}
\author{M. Jonson}
\affiliation{Department of Physics, University of Gothenburg, SE-412 96 G{\" o}teborg, Sweden}
\affiliation{SUPA, Institute of Photonics and Quantum Sciences, Heriot-Watt University, Edinburgh EH14 4AS, Scotland, UK}

\date{\today}
\begin{abstract}
Inefficient screening of electric fields in nanoconductors makes electric manipulation of electronic transport in nanodevices possible. Accordingly, electrostatic (charge) gating is routinely used to affect and control the Coulomb electrostatics and quantum interference in modern nanodevices. Besides their charge, another (quantum mechanical) property of electrons --- their spin --- is at the heart of modern spintronics, a term implying that a number of magnetic and electrical properties of small systems are simultaneously harvested for device applications. In this review the possibility to achieve ``spin-gating" of mesoscopic devices, i.e. the possibility of an external spin control of the electronic properties of nanodevices is discussed. 
Rather than the Coulomb interaction, which is responsible for electric-charge gating, we consider two other mechanisms for spin gating. These are on the one hand the magnetic exchange interaction in magnetic devices and on the other hand the spin-orbit coupling (``Rashba effect"), which is prominent in low dimensional conductors. A number of different phenomena demonstrating the spin gating phenomenon will be discussed, including the spintro-mechanics of magnetic shuttling, Rashba spin splitting, and spin-gated weak superconductivity.\\

\noindent
Keywords: spintronics, spin gating, spin-orbit interaction, magnetic exchange interaction
\end{abstract}

\maketitle

\section{Introduction}
One of the fundamental features of nanoconductors is that their electronic properties quite easily can be influenced by external electric fields. This can achieved by the means of a nearby charged electrode, i.e. by an electrostatic gate, which allows extra electrons to be attracted to or expelled from the nanoconductor. Such an electrostatic gating of nanoconductors is important for the functionality of a number of devices. One example is two-dimensional quantum dot structures, which can be defined in the conducting plane of semiconductor heterostructures by electrostatically confining the lateral orbital motion of the itinerant electrons. \cite{2D.gated.structures} Another example is the single-electron tunneling (SET) transistor, where a gate electrode is used to control the tunneling of single electrons onto its central island and hence the accumulation of charge there. \cite{SET.general.reference} 

The spin degree of freedom of the electrons is essentially decoupled from the orbital electron motion in bulk conductors but may become important for electronic transport in  nanometer sized conductors. This is because the lack of efficient electric screening and the composite nature of certain modern nanodevices opens up a possibility for the spin-orbit coupling to be significant. If so, it becomes possible to control the accumulation of electronic spin in nanoconductors in analogy to how the accumulation of electric charge can be controlled in electrostatically gated nanostructures. As a result ``spin gating" of nanodevices becomes possible, offering new functionality for electronic and spintronic manipulations of nanodevices.

In this article we briefly review some of our recent work that explore the use of spin gating in normal-metal, ferromagnetic and superconducting devices. We will show that non-coherent and quantum-coherent transport in nanowires and nanodots as well as the nanomechanics and superconductance of Josephson weak links can be drastically modified by spin gating, which hence provides a means for spin control on the nanometer length scale. 

The magnetic exchange interaction between the magnetic moments of ferromagnetic leads and the spin accumulated on the central island (quantum dot) is one source of spin gating in magnetic nano-electromechanical SET devices (NEM-SETs). In principle, the exchange force corresponding to this interaction is strong enough to significantly affect the nanomechanics, allowing for spintromechanical polaronic effects and spintromechanical instabilities to appar in such devices.

Another mechanism that can be used for spin gating is the spin-orbit interaction, which can be very strong in nanowires and carbon nanotubes. We will show that this interaction can be employed for spin gating of nanowire-based electric weak links, with the result that incident electronic waves are coherently split with respect to two possible spin projections. This is an effect that can be detected through the spin currents that are generated by such spin-active weak links.

Spin gating of superconducting weak links, finally, induces a spin splitting of the electrons that form Cooper pairs. This may drastically affect the transfer of Cooper pairs through a magnetic wire and opens up the possibility for spin-gate induced stimulation of the Josephson current in Superconductor-Ferromagnet-Superconductor weak links.

The spin-gating phenomena introduced above will be described in more detail in Sections~\ref{Section3} - \ref{Section5} below. Before we do so, we will in Section~\ref{Section2} discuss the physics and the strength of spin-gating effects, which can be expected in modern nanodevices. Finally, In Section~\ref{Conclusions}, we present our conclusions. 

\section{Physics and Strength of Spin-Gating Effects in Modern Nanodevices}
\label{Section2}

\begin{figure}
\vspace{0.0cm} \centerline {\includegraphics[width=9.5cm]{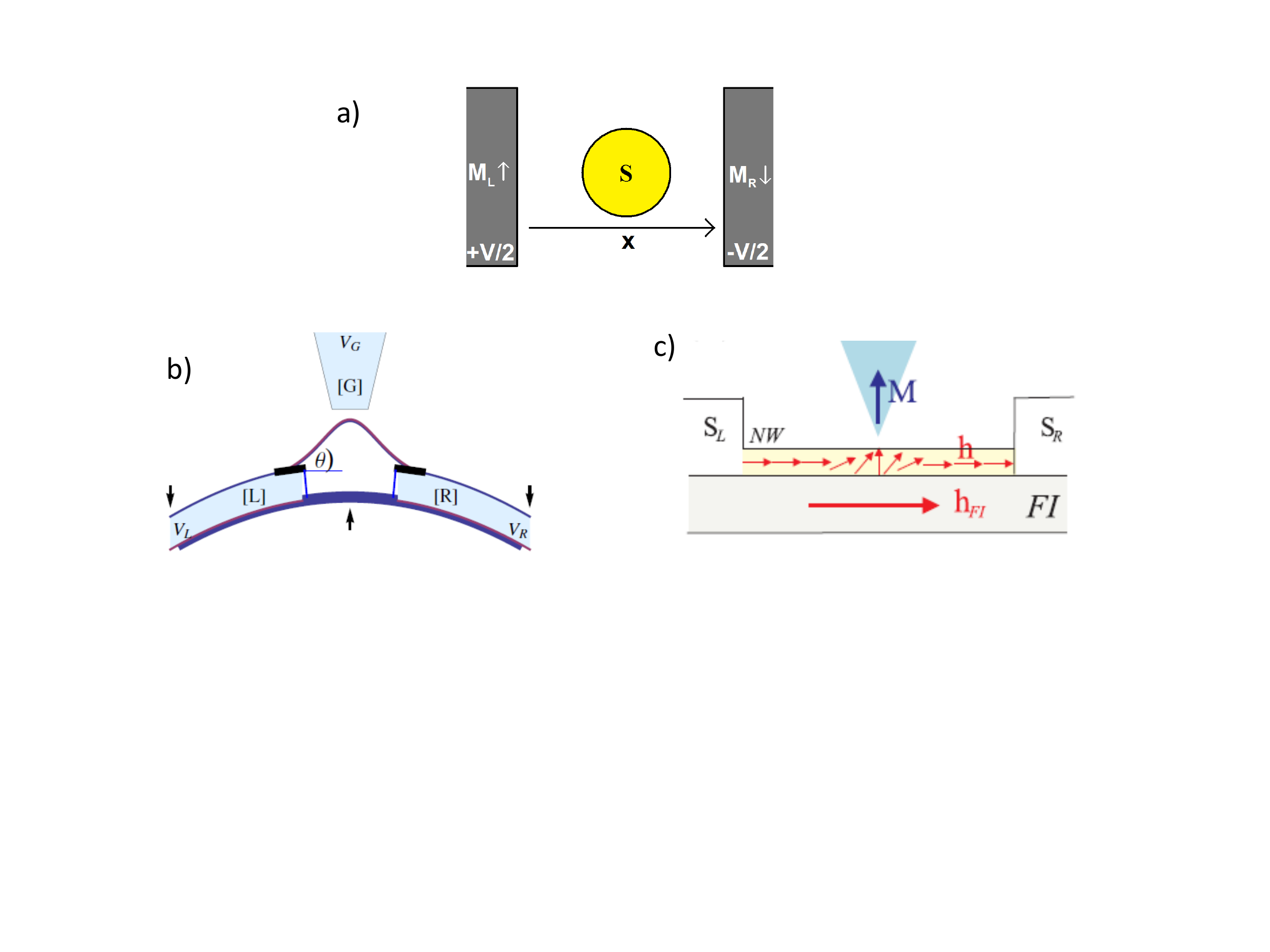}}
\vspace*{-3.1 cm} 
\caption {Sketches of three mesoscopic devices in which a strong spin gating effect can be achieved.
(a) Tunneling device comprising a movable non-magnetic quantum dot coupled by electronic tunneling and by magnetic exchange interactions to ferromagnetic leads. The high sensitivity of the exchange energy to dot displacements along the $x$-axis corresponds to a strong exchange force that affects the nanomechanics of the system.  
(b) The "Rashba spin splitter" discussed in Section~\ref{Section4}: A break junction supports a nanowire 
attached by tunnel contacts to two biased electrodes ([L] and [R]). Small vibrations of the wire induce oscillations in the angle $\theta$ around some value $\theta_0$. 
The upper gate electrode is an STM tip biased differently. The spin-splitting Rashba interaction in the bent wire can be controlled via the bending angle $\theta$, which can be modified both mechanically 
and electrically, by biasing the STM.
(c) The S--F--S constriction discussed in Section~\ref{Section5}: A normal-metal nanowire (NW) in contact with a ferromagnetic insulator (FI) bridges the gap between two superconductors (S) while a magnetic STM tip, acting as a spin gate, can be used to influence the magnetization in parts of the nanowire.
}
\label{Fig2}
\end{figure}

Since there is no problem to allow electric charge to accumulate in nanometer sized spatial domains (nano-dots, micro-constrictions and nano-wires) it is not difficult to produce a strongly localized electrostatic gating effect in nanodevices. In order to achieve a similarly localized spin gating effect by means of an external magnetic field one would need a field that is inhomogeneous on the nanometer scale. This is difficult to achieve by using standard magnetic-field sources. However, a strong spin gating effect can be obtained in a different way in magnetic nanostructures by relying on strong and short range magnetic exchange interactions. A simple nanostructure, where such a magnetic exchange gating can be arranged, is sketched in Fig.~\ref{Fig2}a, which shows a single-electron tunneling device whose source and drain electrodes are ferromagnets. Here, the exchange interaction between the magnetic moment (spin) of electrons localized on the central island of the device (the dot) and the magnetic moments of the electrodes induces a gating effect through the Zeeman energy split of the electronic levels on the dot with respect to their spin projection. The strength of this exchange interaction is proportional to the overlap of the wave functions of electrons localized on the dot with the wave functions of the electrons in the source- and drain electrodes and is therefore exponentially sensitive to the distance between the dot and the electrodes. It follows that the exchange interaction leads to a short-range spin gating effect that varies on the scale of the electron tunneling length. The rapid spatial variation of the exchange energy corresponds to a significant exchange force that acts on the dot and depends on the total amount of spin accumulated in the dot. This is a new feature of magnetic nanodevices compared to nonmagnetic ones, where due to the electrostatic gating effect only the Coulomb force --- proportional to the total electric charge on the dot --- acts on the dot.

The magnetic exchange gating effect described above can be quite strong. In fact the strength of the Zeeman energy split has been measured in several experiments, two of which shall be mentioned here. In one, Pasupathy et al. \cite{Pasupathy} measured Kondo-assisted tunneling via C-60 molecules in contact with ferromagnetic nickel electrodes. Kondo correlations persisted despite the presence of ferromagnetism, but the Kondo peak in the differential conductance was split as a result of an exchange splitting of the Kondo resonance. The local exchange field on the quantum dot was estimated to be greater than 50 tesla. In the other experiment, Hamaya et al. \cite{Hamaya} studied the Kondo effect in a semiconductor quantum dot coupled to ferromagnetic nickel electrodes and found a split Kondo resonance peak in the absence of an external magnetic field. The splitting was again found to be due to a Zeeman spin-split energy level on the dot caused by the magnetic exchange interaction. The splitting of the Kondo resonance could be removed by applying a compensating external magnetic field of about 1.2 tesla, which hence is the approximate strength of the exchange interaction in the experiment. 
 
Another vehicle for coupling electronic spins and the orbital motion of electrons and hence for affecting electronic transport on the nanometer scale is the spin-orbit interaction. \cite{spinorbitintercation} Being relativistically small such a coupling does not play a significant role for transport phenomena in bulk materials but might play an essential role in the vicinity of surfaces, where unscreened electric fields that cause spin-orbit coupling can be as strong as the electric field in atoms. Rashba based his suggestion of a strong spin-orbit effect on the surface properties of solids on this argument. His argumentation is fully applicable to low dimensional conductors such as nanowires and quantum dots with a large surface to volume ratio. In fact the Rashba spin-orbit interaction in carbon nanotubes has been found to significantly affect their electronic properties. 


An important feature of the spin-orbit interaction is its sensitivity to the curvature of the electron trajectory, which is determined by the geometry of nanowire-based nanodevices. This opens the way for inducing an inhomogeneous spin-gating effect in nanowire based electric week links. Later, in Section~\ref{Section4}, we will show that in a mechanically bent nanowire a strong spin gating effect can be induced by the Rashba interaction resulting in a coherent splitting of the electronic states with respect to their spin. Various consequences of such a Rashba spin splitting effect will be discussed. 

Three mesoscopic devices for which strong spin gating effects have been predicted are sketched in Fig.~\ref{Fig2}.  Of these the magnetic shuttle device shown in Fig.~\ref{Fig2}a was used to illustrate spin-gate controlled nanomechanical effects. In this device spin-dependent exchange forces can displace the dot with respect to the source and drain electrodes and thereby influence how electrons of different spin projections tunnel through the device. A strongly spin-polarized electric current and a spin-induced nanomechanical shuttle instability has been predicted for such a ``spintro-mechanical" device. Such spintro-mechanical  phenomena are examples of how spin gating can affect incoherent electron transport through mesoscopic devices. 

The possibility of using spin gating to control the phase of the electronic wave function was demonstrated for the ``Rashba spin-splitter" device shown in Fig.~\ref{Fig2}b. In this case an extra contribution to the phase can be accumulated as electrons travel through a bent nanowire. This effect, known as the Aharonov-Casher effect, can be viewed as a coherent twisting of the electronic spin. It leads to a Rashba spin-gate induced splitting of the electronic waves that travel through the device. Consequences of such a spin-splitting effect for the device functionality are discussed in Section~\ref{Section4}.

How spin-gating of superconducting Cooper pairs can be realized has been demonstrated by considering the Josephson current through a Superconductor-Ferromagnet-Superconductor weak link such as the one sketched in Fig.~\ref{Fig2}c. Here an additional magnetic electrode (spin gate) enables one to control the spin structure of Cooper pairs that travel through the device. In Section~\ref{Section5} we demonstrate the possibility to enhance the supercurrent by such a spin gating technique.
A discussion of future applications of spin gating control of both coherent and incoherent transport phenomena in mesoscopic devices is given in the concluding Section~\ref{Conclusions}.

\section{Spin Gating in Magnetic NEM-SET Devices}
\label{Section3}

\begin{figure}
\vspace{0.cm} \centerline {\includegraphics[width=8cm]{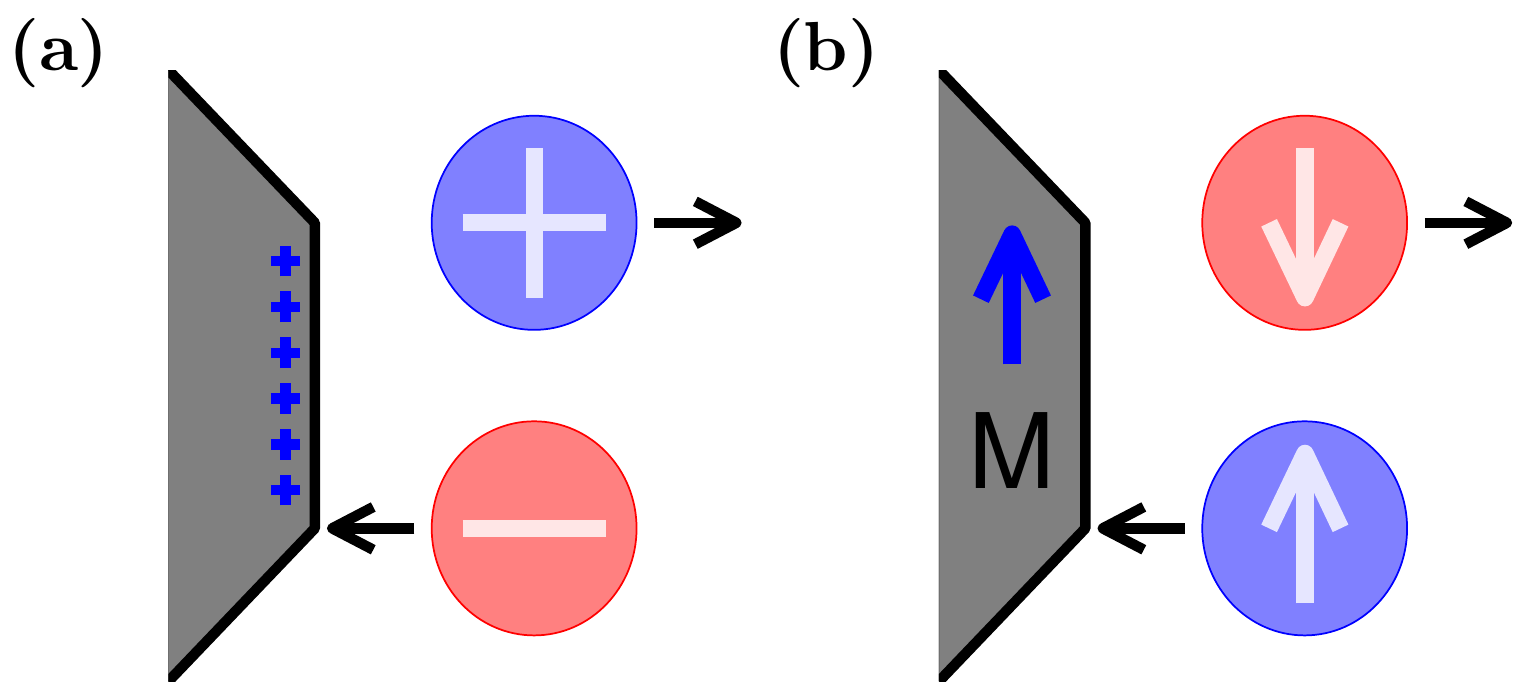}}
\vspace*{-0. cm} \caption{A movable quantum dot in a magnetic
shuttle device can be displaced in response to two types of force:
(a) a long-range electrostatic force causing an electromechanical
response if the dot has a net charge, and (b) a short-rang magnetic
exchange force leading to ``spintromechanical" response if the dot
has a net magnetization (spin). The direction of the force and
displacements depends on the relative signs of the charge and
magnetization, respectively.}
\label{Fig3}
\end{figure}

Electron tunneling from the source- to the drain electrode via the central dot of a single-electron tunneling (SET) device \cite{SET.general.reference} allows for the accumulation of both a net electric charge and a net electron magnetic moment (spin) on the dot. Nanomechanical consequences follow if the dot can change its position as a result of the ensuing and possibly strong electrostatic- and exchange forces as we have discussed in Section~\ref{Section2}. It is interesting to compare the effects of these two forces, which are illustrated in Fig.~\ref{Fig3}. Tunneling of an electron into the dot can be induced by voltage biasing the source electrode. Such a bias results in the accumulation of extra charge on the source electrode so that Coulomb repulsion expels electrons from the electrode and stimulates them to tunnel to the dot. Clearly, the same Coulomb repulsion will then repel the charged dot form the source electrode. 

The situation is qualitatively different if extra spins are accumulated on a magnetic source electrode so that a non-zero net magnetization of the material is possible. In this case majority-spin electrons have a lower energy due to their exchange interaction with the magnetization of the material. Making such electrons tunnel to the dot diminishes this energy lowering effect, which can be viewed as an increase of the exchange energy. It follows that an exchange force resulting from localizing a majority-spin electron on the dot would attract the dot towards the source electrode. This means that electrostatic biasing and spin biasing of electrons on the dot induce forces on the dot with opposite directions. The result is a qualitative difference between electrostatically gated electromechanics \cite{electrostaticallygatedmechanics} and spin-gated spintro-mechanics, the latter being the subject of this Section. 

Another difference between electro- and spintro-mechanical behavior is that in contrast to electric charge the electronic spin accumulated on the dot can be changed even in the absence of electron tunneling into and out of the dot. This can be done by applying an external magnetic field applied perpendicular to the spin polarization of the electrons. The effect of such magnetic stimulation of mechanical dot vibrations is discussed below.

An interesting effect of spatial spin segregation in a NEM-SET device occurs if the spin-dependent shift of the equilibrium position of its central dot is taken into account. This shift happens since in addition to the elastic forces there is a magnetic exchange force whose sign depends on the net magnetic moment of the dot (which could be due to a single electron spin). The equilibrium dot position is shifted either closer to or farther away from the source electrode and hence either farther away from the drain electrode or closer to it. As a result the equilibrium position of the dot will be different depending on the orientation of the electron spin(s) on the dot. Since the tunneling resistance is exponentially sensitive to the position of the dot relative to the leads it follows that tunneling currents comprising electrons of different net spin orientations could be of very different magnitude. That the effect of such spin filtering can be very large will be demonstrated below.

The Hamiltonian that describes the magnetic nanomechanical SET device has the standard form, except for its spin-dependent part (representing the magnetic exchange energy) which now depends on the mechanical displacement of the dot. Hence
\begin{equation}
H = H_{\rm leads} + H_{\rm tunnel} + H_{\rm dot}\,,
\label{eq:hamiltonian}
\end{equation}
where
\begin{equation}
H_{\rm leads} = \sum\limits_{\mathbf{k}, \sigma, s} a_{\mathbf{k} s \sigma}^\dagger a_{\mathbf{k} s \sigma} \epsilon_{\mathbf{k}s \sigma}
\end{equation}
describes electrons  (labeled by wave vector $\mathbf{k}$ and spin $\sigma=\uparrow, \downarrow$) in the two leads ($s=L$, $R$). 
Electron tunneling between the leads and the dot is modeled as  
\begin{equation}
H_{\rm tunnel} = \sum\limits_{\mathbf{k},\sigma, s}T_s \left ( x \right ) a_{\mathbf{k} s \sigma}^\dagger c_{\sigma} + H.c.,
\label{eq:htunnel}
\end{equation}
where the matrix elements $T_{s} (x)= T_{s}^{(0)} \exp(\mp x/\lambda)$, with $\lambda$ the characteristic tunneling length,
depend on the dot position $x$.

The movable single-level dot is modeled as a harmonic oscillator of angular frequency $\omega_0$,
\begin{equation}
\begin{array}{c}
H_{\rm dot} = \hbar \omega_0 b^\dagger b + \sum\limits_{\sigma} n_\sigma \left[ \epsilon_0 - {\rm sign}(\sigma) J \left ( x \right ) \right] + E_C n_{\uparrow} n_{\downarrow}\,,
\end{array}
\label{eq:hgrain}
\end{equation}
where ${\rm sign}(\uparrow,\downarrow)=\pm 1$, $E_C$ is the Coulomb energy associated with double occupancy of the dot and the eigenvalues of the electron number operators $n_\sigma$ is 0 or 1. The position dependent magnitude $J(x)$ of the spin dependent shift of the electronic energy level on the dot is due to the exchange interaction with the magnetic leads (and any external magnetic field).

Tunneling of electrons into or out of the dot may have two distinct effects on the dot mechanics. One is that it in principle changes the equilibrium position of the dot with respect to the leads (a polaronic effect) and the other is that it may lead to mechanical vibrations of the dot. The build-up of the amplitude of these vibrations as more and more electrons pass through the device could --- depending on the strength of the mechanical dissipation in the vibronic subsystem --- lead to a steady-state amplitude much larger than in thermal equilibrium. Below we will consider the two opposite limits of strong and weak dissipation. In the strong-dissipation limit no non-equilibrium vibrations will develop and small-amplitude vibrations of the dot will correspond to thermal equilibrium. In the  weak-dissipation limit, on the other hand, the amplitude of the dot vibrations may be much larger than in thermal equilibrium and correspond to a current-induced nanomechanical shuttle instability.

\subsection{Spin-Polaronic Discrimination of Spin-Polarized Electrical Currents}
\label{Section3A}

\begin{figure}
\vspace{0.cm} \centerline {\includegraphics[width=6cm]{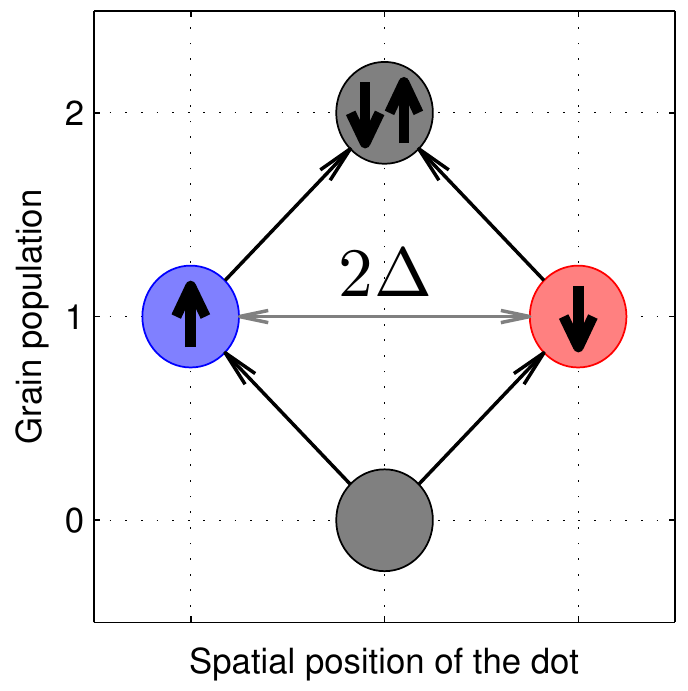}}
\vspace*{-0. cm} \caption{Diagram showing how the equilibrium
position of the movable dot depends on its net charge and spin. The
difference in spatial displacements discriminates transport through
a singly occupied dot with respect to the electron spin.}
\label{Fig4}
\end{figure}

The spatial separation of dots with opposite spins
is illustrated in Fig.~\ref{Fig4}. While changing the population of spin-up
and spin-down levels on the dot (by changing e.g. the bias voltage
applied to the device) one shifts the spatial position $x$ of the
dot with respect to the source/drain leads. It is important that the
Coulomb blockade phenomenon prevents simultaneous population of both
spin states.
If the Coulomb blockade
is lifted the two spin states become equally populated with a zero
net spin on the dot.
This removes the spin-polaronic
deformation and the dot is situated at the same place as a
non-populated one. In calculations a strong modification of the
vibrational states of the dot, which has to do with a shift of its
equilibrium position, should be taken into account. This results in
a so-called Frank-Condon blockade of electronic tunneling. \cite{116,
ratner} The spintro-mechanical stimulation of a spin-polarized
current and the spin-polaronic Franck-Condon blockade of electronic
tunneling are in competition and their interplay determines a
non-monotonic voltage dependence of the giant spin-filtering effect.

\begin{figure}
\vspace{0.cm} \centerline {\includegraphics[width=8cm]{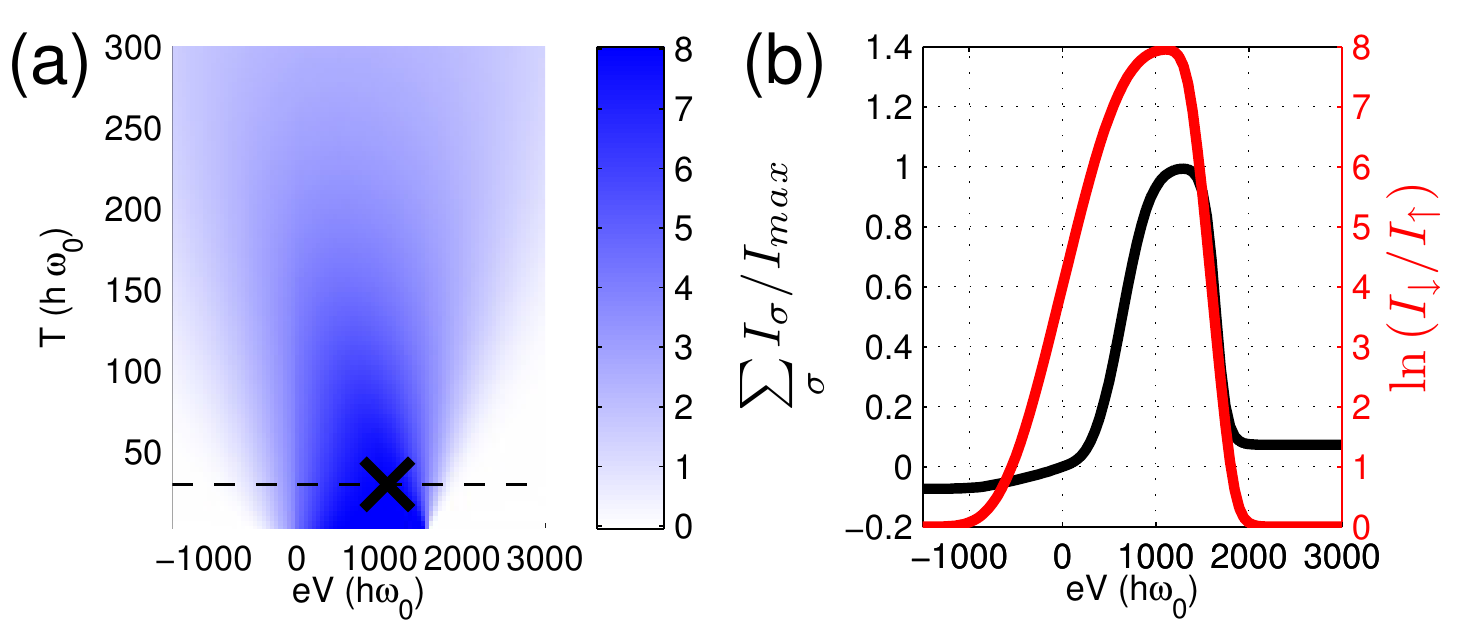}}
\vspace*{-0. cm} \caption{ Spin polarization of the current through
the model NEM-SET device under discussion.}
\label{Fig5}
\end{figure}

To understand the above effects in more detail consider the
analytical results of Ref.~\onlinecite{7}. A solution of the problem can
be obtained by the standard sequential tunneling approximation and
by solving a Liouville equation for the density matrix for both the
electronic and vibronic subsystems. The spin-up and spin-down
currents can be expressed in terms of the tunneling rates $\Gamma_{L,R}$ (energy
broadening of the level) and the occupation probabilities for
the dot electronic states.
For simplicity we consider the case of a strongly asymmetric
tunneling device. At low bias voltage and low temperature the
partial spin current is
\begin{equation} \label{Spint curr1}
I_{\sigma}\sim 
e \Gamma_L \exp\left(\frac{1}{2}\left[\frac{a_0^2}{\lambda^2}-\left(\frac{a_0}
{\hbar\omega_0}\right)^2\right]-{\rm sgn}(\sigma)\beta\right),
\end{equation}
where $\beta = \Delta/\lambda$ is the ratio of the polaronic shift 
$\Delta$
of the equilibrium spatial position
of a spin-polarized dot and the electronic tunneling length $\lambda$
($\Delta = \vert J'(0) \vert a_0^2/\hbar \omega_0$ where $a_0$ is the zero-point oscillation amplitude of the dot).
 In the high bias voltage
(or temperature) regime, ${\rm max}\{eV,T\} \gg E_p=\alpha_p^2\hbar\omega_0$ ($\alpha_p=\Delta/a_0$), where the polaronic
blockade is lifted (but double occupancy of the dot is still
prevented by the Coulomb blockade), the current expression takes the
form
\begin{equation} \label{Spintr curr2}
I_{\sigma}\sim 
e \Gamma_L\exp\left(\left[2n_B+1\right]\frac{a_0^2}{\lambda^2}-2\,
{\rm sgn}(\sigma)\beta\right),
\end{equation}
where $n_B$ is Bose-Einstein distribution function. The scale of the
polaronic spin-filtering of the device is determined by the parameter
$\beta$,
which for typical values of the exchange interaction and mechanical properties
of suspended carbon nanotubes 
is about 1-10. As was
shown this is enough for the spin filtering of the electrical
current through the device to be nearly 100 \% efficient. The
temperature and voltage dependence of the spin-filtering effect is
presented in Fig.~\ref{Fig5}. The spin filtering effect and the Franck-Condon
blockade both occur at low voltages and temperatures (on the scale of the
polaronic energy; see Fig.~\ref{Fig5}a).  An increase of the voltage
applied to the device lifts the Franck-Condon blockade, which
results in an exponential increase of both the current and the
spin-filtering efficiency of the device. This increase is blocked
abruptly at voltages for which the Coulomb blockade is lifted. At
this point a double occupation of the dot results in spin
cancellation and removal of the spin-polaronic segregation. This
leads to an exponential drop of both the total current and the spin
polarization of the tunnel current (Fig.~\ref{Fig5}b). As one can see in
Fig.~\ref{Fig5}b prominent spin filtering can be achieved for realistic device
parameters. The temperature of operation of the spin-filtering
device is restricted from above by the Coulomb blockade energy. One
may, however, consider using functionalized nanotubes
\cite{pulkin20} or graphene ribbons \cite{pulkin21} with one or more
nanometer-sized metal or semiconductor nanocrystal attached. This
may provide a Coulomb blockade energy up to a few hundred kelvin,
making spin filtering a high temperature effect. \cite{7}

\subsection{Spintro-mechanical shuttling of electrons}
\label{Section3B}

\begin{figure}\vspace{0.cm} \centerline {\includegraphics[width=9.8cm]{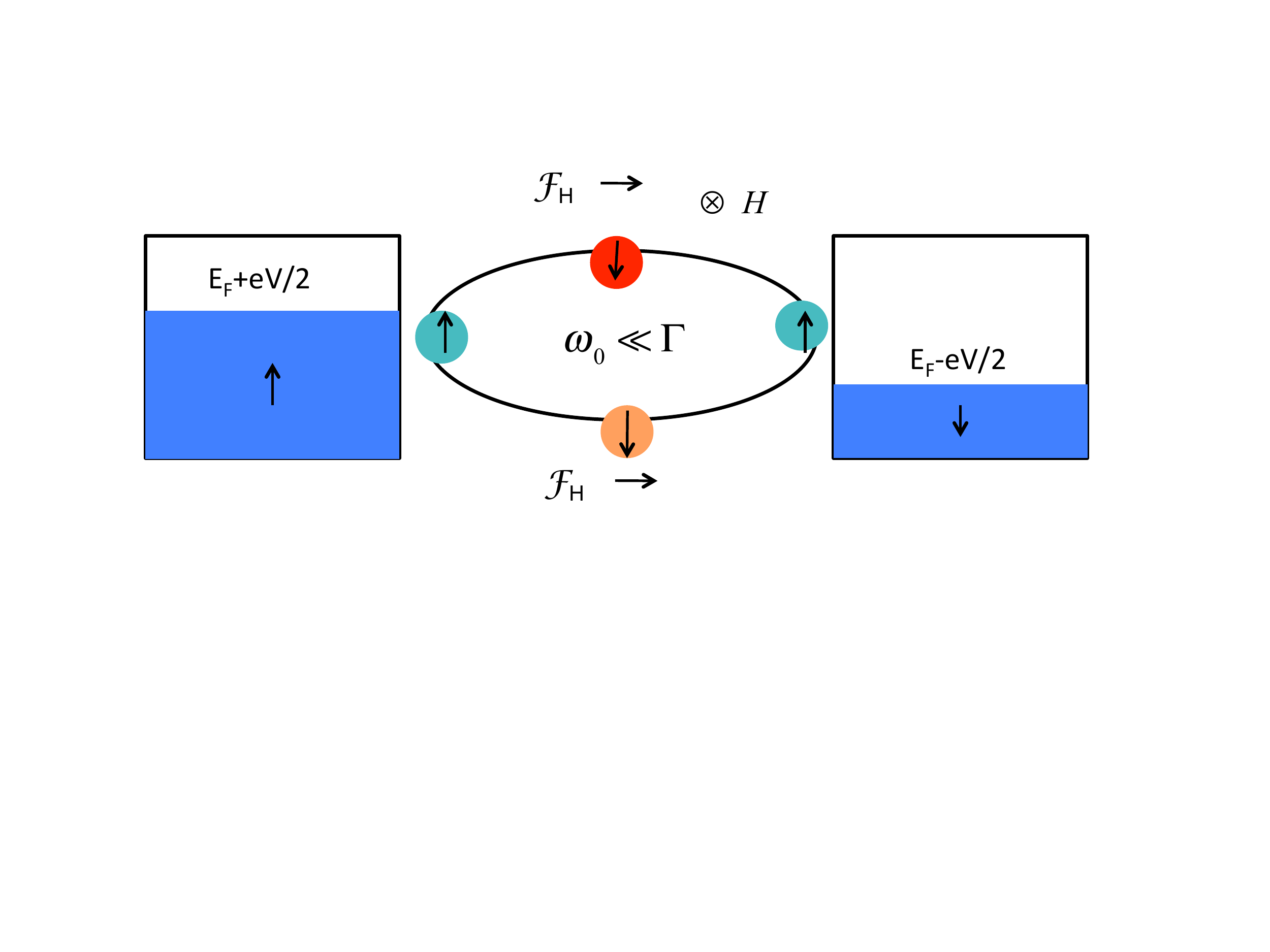}}
\vspace*{-3.5 cm} \caption{Sketch of a voltage biased device where a movable dot may oscillate between two fully spin polarized leads with anti-parallel magnetization directions in the presence of a perpendicular magnetic field $H$. If $H=0$ no current can flow between the leads due to a so called spin blockade. Note that a spin-up electron on the dot is attracted to the source and repelled from the drain by the magnetic exchange force. However, the resulting force does not do any work during a complete oscillation cycle since the spin in this case is a constant of motion. If $H \ne 0$ spin flips may occur, which opens up the possibility for a current to flow. A flipped spin on the dot will be attracted to the drain and repelled from the source by a force ${\cal F}_H(x)$. Averaging over the position $x$ for spin flips one finds that this force may do positive work on the dot and hence lead to a spintro-mechanical shuttle instability. 
}
\label{Fig6}
\end{figure}

In this Subsection we will focus on weakly dissipating nanomechanical shuttle devices. In that case the coupling of the vibrations to a non-equilibrium flux of electrons may drive the mechanical subsystem far away from thermal equilibrium. This is what happens if a finite energy transfer between electrons and vibrons results in energy being pumped into nanomechanical vibrations. It is illustrative to analyze the criterion for such a pumping to occur by considering the work done by the spintromechanical force associated with electron tunneling. For this purpose we adopt a simple model and consider a movable quantum dot comprising one spin up and one spin down electronic state. The dot can oscillate around an equilibrium position between two fully spin polarized ferromagnetic leads with anti-parallel magnetization directions, as sketched in Fig.~\ref{Fig6}.

If tunneling of electrons is prohibited and there is no external magnetic field, then the total electronic spin located on the dot is conserved. It follows that no net work is done by the spintro-mechanical force during one oscillation period and no change of the dot vibration energy is possible due to this force. Now, what would be effect of electronic tunneling? 

The main effect of tunneling is to provide a conductance mechanism by which electrons can tunnel from the source electrode to the dot and then from the dot to the drain electrode. Tunneling of an electron from the source to the dot changes the total value of its spin by the spin of the tunneling electron and hence adds to the spintro-mechanical force on the dot. Could this additional force do work on the dot as it carries the extra electron towards the drain electrode? The answer is no! The reason is that the spin projection of the extra electron is the same as that of the majority spins in the source and that therefore the extra force is a retardation force directed opposite to the velocity of the dot.
The conclusion is that extra mechanical dissipation is created spintro-mechanically and that no pumping of vibrational energy is possible in this case. 

However, as we have already pointed out, the situation is radically changed if an external magnetic field oriented perpendicular to the spin is switched on. Such a magnetic field induces electronic spin flips, which reverse the direction of the additional spintro-mechanical force. As we will show in this Subsection a perpendicular magnetic field under certain conditions facilitates nanomechanical vibrations and a spintro-mechanical instability, which can develop into self-sustained, large-amplitude dot vibrations. Such vibrations, which are accompanied by mechanical transportation of both electronic charge and net electronic spin, we call spintro-mechanical shuttling while we refer to the instability that starts the shuttling as a spintro-mechanical shuttle instability.



A particularly transparent picture of how spintro-mechanics affect
shuttle vibrations emerges in the limit of weak magnetic field $H$
and large electron tunnelling rate $\Gamma_{S(D)}$ between dot and
source- and drain electrodes. In order to explore this limit,
where $\Gamma_{S}\gg\omega_0\gg(\mu H/\hbar)^{2}/\Gamma_{D}$ and
$\omega_0/2\pi$ is the natural vibration frequency of the dot, we
focus first on the total work done by the exchange force
$\mathfrak F$ as the dot vibrates under the influence of an
elastic force only. In the absence of an external magnetic field
\cite{note1} the dot is in this case occupied by a spin-up
electron emanating from the source electrode. 
This spin is a constant of motion and hence no
electrical current through the device is possible since only
spin-down states are available in the drain electrode. During the
oscillatory motion of the dot the exchange force is therefore
always directed towards the source electrode while its magnitude
only depends on the position of the dot, $\mathfrak F=\mathfrak
F_0(x)$. As a result, no net work is done by the exchange force on
the dot. This is because contributions are positive or negative
depending on the direction of the dot's motion and cancel when
summed over one oscillation period. A finite amount of work can
only be done if the exchange force deviates from $\mathfrak
F_0(x)$ as a result of spin flip processes induced by the external
magnetic field. Such a deviation can be viewed as an additional
random force $\mathfrak F_H$ that acts in the opposite direction
to $\mathfrak F_0(x)$. In the limit of large tunneling rate,
$\Gamma_{S(D)}\gg\mu H/\hbar$, and small vibration amplitude a
spin flip occurs  with a probability $\propto(\mu
H/\hbar)^2/(\omega_0\Gamma_D)$ during one oscillation period and is
instantly \cite{instant} accompanied by the tunneling of the dot
electron into the drain electrode, thereby triggering the force
$\mathfrak F_H$. The duration of this force is determined by the
time $\delta t\sim 1/\Gamma_S(x(t))$ it takes for the spin of the
dot to be ``restored" by another electron tunneling from the
source electrode.

The spin-flip induced random force $\mathfrak F_H=-\mathfrak
F_0(x)$  is always directed towards the drain electrode. Hence,
its effect depends on the dot's direction of motion: as the dot
moves away from the source electrode it will be accelerated, while
as it moves towards the source it will be decelerated. Since a
spin-flip may occur at any point on the trajectory one needs to
average over different spin-flip positions in order to calculate
the net work done on the dot. The result, which depends on the
competition between the effect of spin flips that occur at the
same position but with the dot moving in opposite directions, is
nonzero because $\delta t$ is different in the two cases. As the
dot moves away from the source electrode the tunneling rate to
this electrode will decrease while as the dot moves towards the
source it will increase. This means that the duration of spin-flip
induced acceleration will prevail  over the one for deceleration.
As a result, in weak magnetic fields, the dot will accelerate with
time and one can expect a spintro-mechanical  shuttle instability
in this limit.

The situation is qualitatively different in the opposite limit of
strong magnetic fields, where $\Gamma_{S(D)}\ll\mu H/\hbar$ and
the spin rotation frequency therefore greatly exceeds the
tunneling rates. In this case the quick precession of the electron
spin in the dot averages the exchange force to zero if one
neglects the small effects of electron tunneling to and from the
dot. If one takes corrections due to tunnelling into account
(having in mind that the source electrode only supplies spin-up
electrons) one comes to the conclusion that the average spin on
the dot will be directed upwards. This results in a net
spintro-mechanical force in the direction opposite to that of the
net force occurring in a weak magnetic field limit. As a result,
in strong magnetic fields one expects on the average a
deceleration of the dot. Therefore, there will be no shuttle
instability for such magnetic fields.

As we have discussed above spin-flip assisted electron tunnelling
from source to dot to drain in our device results in a magnetic
exchange force that attracts the dot to the source electrode. It
is interesting to note that this is contrary to the effect of the
Coulomb force in the same device. \cite{rev2comment} Indeed, since
the Coulomb force depends on the electric charge of the dot it
repels the dot from the source electrode. Hence, while the dot is
empty as the result of a spin-flip assisted tunneling event from
dot to drain, an ``extra" {\em attractive} Coulomb force
$\mathfrak F_Q$ is active. An analysis fully analogous with our
previous analysis of the ``extra" {\em repulsive} magnetic
exchange force $\mathfrak F_H$  leads to the conclusion that the
effect of the Coulomb force will be just the opposite to that of
the exchange force. If the exchange force is sufficiently weak,
this means that in the Coulomb blockade regime there is no shuttle
instability in the limit of weak magnetic field, while in strong
magnetic fields electron shuttling occurs. A full analysis, which confirms 
the predictions made above for some limiting cases using
only qualitative arguments, can be found in
Ref.~\onlinecite{ourPRL2014}.

\section{Spin Gating of Quantum Coherent Electron Transport Through Nanowire-Based Electric Weak Links}
\label{Section4}

Quantum coherence did not play any role in the spin-gate induced nanomechanical phenomena considered so far in this review. From now on, however, we will focus on this very subject. i.e. on spin-gated phase coherent transport. 

Spin gating of the phase of the electronic wave function can be achieved via the Aharonov-Casher effect, \cite{AharonovCasher} which is that if an electron spin moves in an external electric field an extra phase is accumulated as the electron moves along its trajectory. 
The effect is a direct consequence of classical electrodynamics being applied to the relativistically small magnetic moment induced by the spin of an electron. It is dual to the Aharonov-Bohm effect, where an extra phase is accumulated due to propagation of an electron along its trajectory in an external {\em magnetic} field.

Classically, the effect of the interaction between a magnetic moment and an electric field is readily obtained by observing \cite{EurphysicsJounal1991} that a magnetic moment $\vec{\mu}$ moving with velocity ${\bf v}$ gives rise to a an electric dipole moment
\begin{equation}
{\bf P}  = \frac{1}{c} \left[{\bf v} \times \vec{\mu} \right]
\end{equation}
in the rest frame of the charges responsible for the electric field ${\bf E}$. The interaction energy is then given by \cite{alternative}
\begin{equation}
\label{U}
U = - {\bf P}\cdot {\bf E} =  - \frac{1}{2mc} \,\vec{\mu} \cdot \left( {\bf E} \times {\bf p} \right) \,.
\end{equation}
Equation~(\ref{U}) gives the correct value \cite{EurphysicsJounal1991} for the spin-orbit interaction of an electron if one uses the Bohr magneton $\mu_B=e\hbar/(2mc)$ for the magnetic moment associated with the electronic spin.  Being a relativistic effect, the electron spin-orbit coupling is generically very weak and a large electric field is therefore required for it to have any significant strength. This is hard to achieve in bulk metals, where the screening of electric fields is very efficient. However, as was pointed out by Rashba, \cite{Rashba} strong enough electric fields may appear in the vicinity of a crystal surface, where the unscreened electric field induced by surface band bending can be as strong as the electric field in heavy atoms. Such an enhanced spin-orbit interaction, now called Rashba spin-orbit interaction, can dominate the electronic properties of low dimensional conductors such as quantum dots and nanowires, which serve as electric weak links in mesoscopic devices. 

In order to show that a significant spin gating effect can be induced by Rashba spin-orbit coupling, 
consider the mesoscopic device shown in Fig.~\ref{Fig2}b. Here a bent nanowire serves as an electric weak link between two bulk electron reservoirs. If an electron enters the wire from the left reservoir, say, its kinetic energy has to change. This is because of the finite spin-orbit interaction in the wire and the condition that the total energy of the electron has to be conserved. The change in kinetic energy corresponds to a change of momentum and  hence of the de Broglie wave length of the electron. Accordingly the electron wave function will accumulate an extra phase as the electron travels along its trajectory through the wire,
\begin{equation}
\label{extraphase}
\psi({\bf r}) \sim \exp\left(\frac{i}{2c\hbar}\int \left( {\bf \mu} \times {\bf E} \right)\cdot d{\bf r} \right)\,.
\end{equation} 
This is nothing but a manifestation of the Aharonov-Casher effect and represents a spin gating effect on coherent electron transport. The effect can be incorporated as an extra phase factor in an effective probability amplitude for electron tunneling through a nanowire-based tunneling weak link such as that shown Fig.~\ref{Fig2}b. 

\begin{figure}\vspace{0.cm} \centerline {\includegraphics[width=9cm]{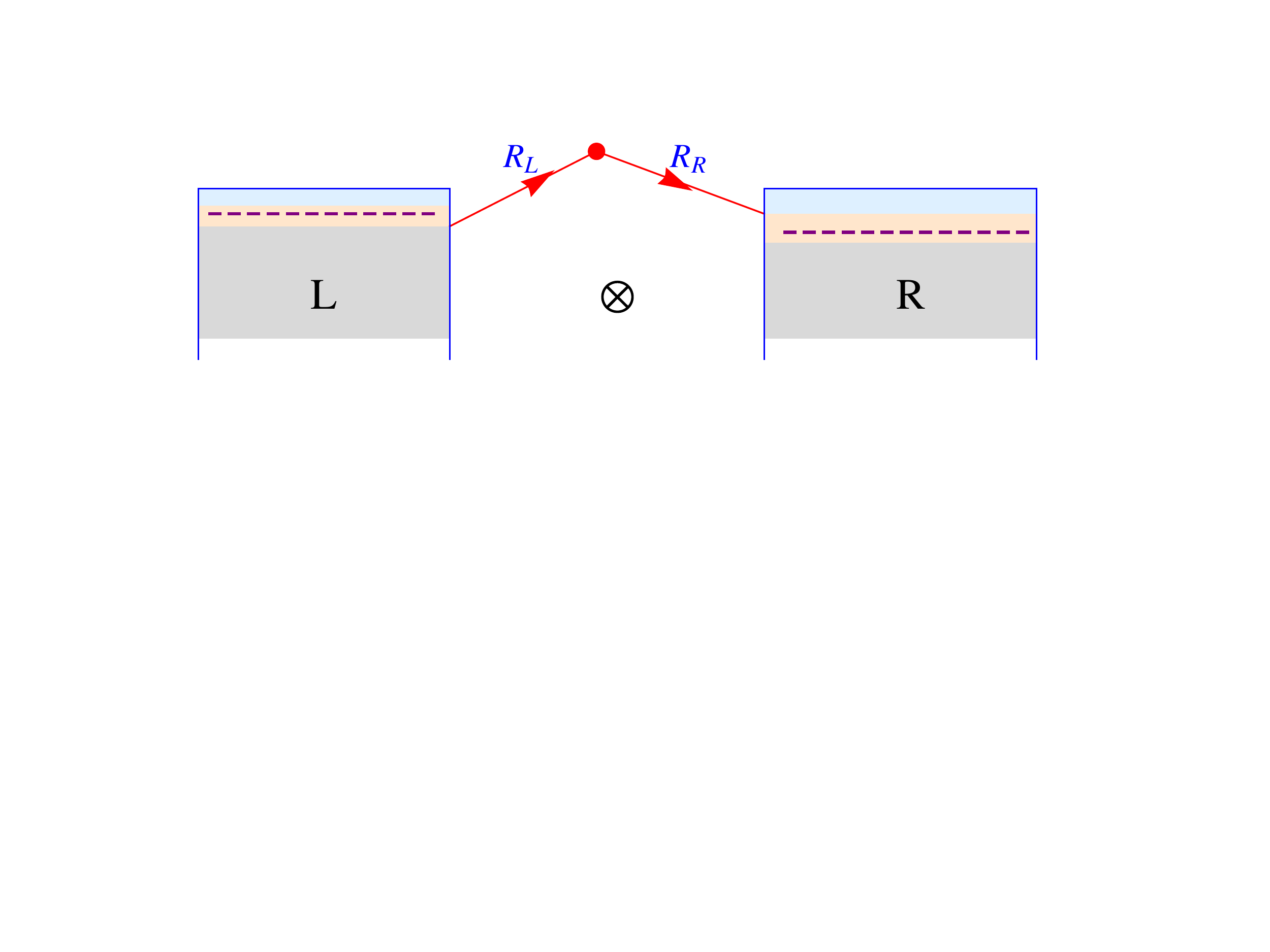}}
\vspace*{-4.0 cm} 
\caption{Schematic geometry used for calculating the spin-orbit coupling dependence of the tunneling 
amplitude for the device shown in Fig.~\ref{Fig2}b. A localized level is tunnel coupled to left ($L$) and 
right ($R$) electrodes with possibly different chemical potentials $\mu_{L\sigma}$ and 
$\mu_{R\sigma}$. The setup lies in the $x$-$y$ plane and a magnetic field is applied along $\hat z$. 
It corresponds to a configuration in which the wire is controlled only mechanically, and the 
STM is not shown.}
\label{Fig7}
\end{figure}

It is convenient to use the simplified model system shown in  Fig.~\ref{Fig7} for calculations (see Ref.~\onlinecite{Ora.Amnon.RS.PRL} for details). There, the nanowire is replaced by a quantum dot,
which has a single level (of energy $\epsilon_0$), and which vibrates in the direction perpendicular to the wire in the junction plane. The leads are modeled by free electron gases and are firmly coupled to left and right reservoirs, of chemical potentials $\mu_{L\sigma}$ and $\mu_{R\sigma}$, respectively, allowing for spin-polarized charge carriers. Here, $\sigma$ denotes the spin index; the spin-quantization axis (assumed to be the same for both reservoirs) depends on the spin imbalance in the reservoirs and will be specified below. The electronic populations in the reservoirs are thus
\begin{equation}
f_{L(R)\sigma}(\epsilon_{k(p)}) =1/ \left[e^{\beta(\epsilon_{k(p)}-\mu_{L(R)\sigma})} + 1 \right]
\end{equation}
where $\beta = 1/k_BT$. 
The electron gas states in the left (right) lead are indexed by $k$ ($p$) and have energies $\epsilon_k$ ($\epsilon_p$). Below we denote by $c_{k\sigma}$ ($c_{p\sigma}$) the annihilation operators for the leads, and by $c_{0\sigma}$ that for the localized level.

The linear Rashba interaction manifests itself as a phase factor on the tunneling amplitude. \cite{Ora2005} In the geometry of Fig.~\ref{Fig7}, this phase is induced by an electric field perpendicular to the $x$-$y$ plane, and is given by $\alpha_{\rm so} \, {\bf R} \times \vec{\sigma} \cdot \hat z$, where $\alpha_{\rm so}$ denotes the strength of the spin-orbit interaction (in units of inverse length), and $\vec{\sigma}$ is a vector whose components are the Pauli matrices. Quite generally, ${\bf R}_L = (x_L, y_L)$ for the left tunnel coupling and ${\bf R}_R = (x_R, -y_R)$ for the right one, where both radius vectors ${\bf R}_L$ and ${\bf R}_R$ are functions of the vibrational degrees of freedom (as specified in the following). The quantum vibrations of the wire, which modify the bending angle, make the electronic motion effectively two dimensional. This leads to the possibility of manipulating the junction via the Aharonov-Bohm effect, by applying a magnetic field which imposes a further phase on the tunneling amplitudes $\phi_{L(R)} = - (\pi/\Phi_0)(Hx_{L(R)}y_{L(R)})$, where $H$ is the magnetic field and $\Phi_0$ is the flux quantum (a factor of order unity is absorbed \cite{ShekhterPRL2006} in $H$ ).

It follows that the tunneling Hamiltonian between the localized level and the leads takes the form
\begin{eqnarray}
H_{\rm tun}&=&\sum_{k,\sigma,\sigma'} \left( V_{k\sigma\sigma'} c_{0\sigma}^{\dag}c_{k\sigma'} + H.c. \right) \\
&& + \sum_{p,\sigma,\sigma'} \left( V_{p\sigma\sigma'} c_{p\sigma}^{\dag}c_{0\sigma'} + H.c. \right)\,. \nonumber
\end{eqnarray}
The tunneling amplitudes are (operators in spin and vibration spaces) 
\begin{equation}
V_{k(p)} = - J_{L(R)}\exp\left(-i\psi_{L(R)}\right)\,, 
\end{equation}
where
\begin{eqnarray}
\psi_L &=& \phi_L - \alpha_{\rm so}\left(x_L\sigma_y - y_L\sigma_x\right)\,, \\
\psi_R &=& \phi_R - \alpha_{\rm so}\left(x_R\sigma_y + y_R\sigma_x\right)\,. \nonumber
\end{eqnarray}
We consider a non-resonant case, where the localized level is far above the energies of the occupied states in both leads (i.e., no energy level in the wire is close enough to $\epsilon_0$ for it to be involved in inelastic tunneling via a real state). This allows us to exploit the tunneling as an expansion parameter  \cite{ShekhterPRL2006} and to preform a unitary transformation which replaces the wire by an effective direct tunneling between the leads through virtual states
\begin{equation}
H_{\rm tun}^e = \sum_{k,p} \left( c_k^\dag W_{kp}^\dag c_p + H.c. \right)
\end{equation}
with (using matrix notations in spin space)
\begin{equation}
W_{kp}^\dag = \frac{1}{2}\left( \frac{1}{\epsilon_p - \epsilon_0} + \frac{1}{\epsilon_k - \epsilon_0} \right)V_k^\dag V_p^\dag
\end{equation}

A straightforward calculation \cite{Ora.Amnon.RS.PRL} now gives the spin currents in the system, in particular the currents from the weak link into the left and right leads. These currents may carry current and/or spin and will obviously depend on the chemical potentials in the two leads, which in general can be written as $\mu_{L,R,\uparrow} = \mu_{L,R}+U_{L,R}/2$ and $\mu_{L,R,\downarrow} = \mu_{L,R}-U_{L,R}/2$.
A particularly interesting case is when $\mu_L = \mu_R$ and $U_L = U_R = U \ne 0$, so that there is no electrical bias but only  a spin bias.
Such a spin bias can be achieved by pumping spin into the bulk electrodes either by injecting a spin polarized current or by irradiation with circularly polarized light producing photon induced electronic spin-flip transitions. In this case a finite spin current $J_{\rm spin,\uparrow} = - J_{\rm spin,\downarrow}$ is generated inside the Rashba weak link and pumped out into the leads in such a way as to counteract the spin bias (which we assume to be maintained by external pumping as described above). 
Introducing a spin conductance $G_{\rm spin}$ in the linear response regime, where $J_{{\rm spin}, \uparrow} = -U G_{\rm spin}$, one finds the following results for $G_{\rm spin}$ in the high- and low temperature limits:
\begin{equation}
\frac{G_{\rm spin}}{G_0} = \sin^2(\alpha_{\rm so} d) \cos^2(\theta_0) \times 
\left\{ \begin{array} {ll} 1-\frac{\beta\hbar\omega_0}{6}\frac{H^2}{H_0^2} & \beta\hbar\omega_0 \ll 1 \\ e^{-H^2/H_0^2} & \beta\hbar\omega_0 \gg 1 \end{array} \right.
\end{equation}
Here $H_0 = \sqrt{2}\Phi_0/\left[\pi d a_0 \cos(\theta_0) \cos(2\theta_0)\right]$ gives the magnetic field scale. For values of $\alpha_{\rm so}$ typical for experiments and a nanowire of length $d \sim 1$~$\mu$m one finds that the spin conductance is of the same order as the usual (electrical) conductance $G_0$ (divided by $e^2$); $G_0=\hbar \Gamma_L \Gamma_R/(\pi\epsilon_0^2)$.

In addition to spin current injection, other transport phenomena in a Rashba spin splitter can be considered, including electric and thermal transport through the Rashba spin splitter connected by both magnetic and non-magnetic leads. A detailed description of a number of such phenomena, including a spin-gate controlled photovoltaic effect, can be found in Ref.~\onlinecite{Ora.Amnon.RS.PRB}.

\section{Spin Gating of Cooper Pairs in Superconducting Weak Links}
\label{Section5}

Weak superconductivity is a phenomenon that relies on the so called proximity effect at the boundary between a superconductor and a normal metal. As Cooper pairs of electrons are injected from the superconductor into the normal metal the correlation between the members of the pairs persists for a while in the proximity of the superconductor. If the distance between two such boundaries, forming a superconductor--normal metal--superconductor weak link, does not exceed the superconducting coherence length the proximity effect allows a superconducting current to flow through the device. 

\begin{figure}
\vspace{0.cm} \centerline {\includegraphics[width=9cm]{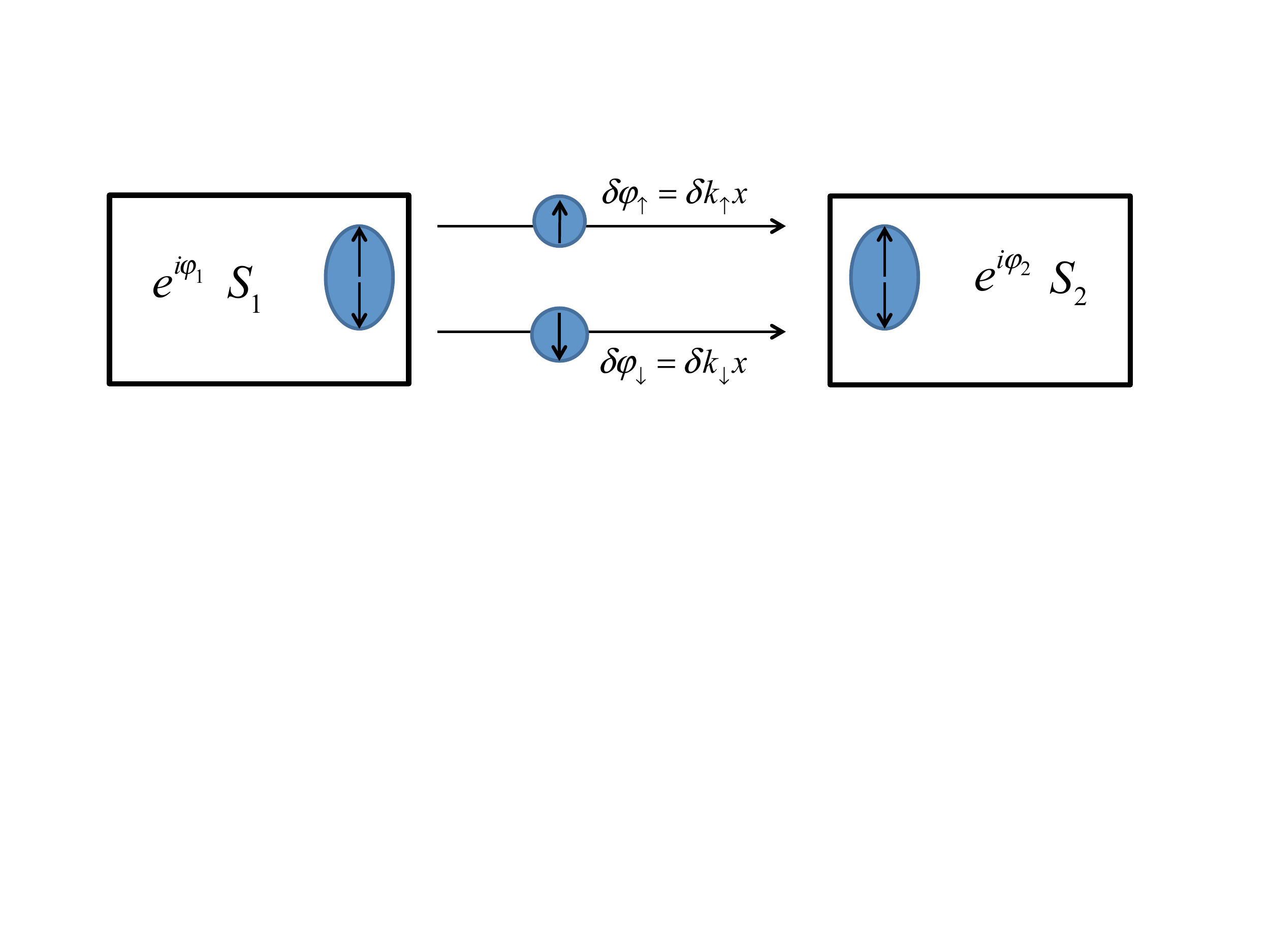}}
\vspace*{-4.0 cm} 
\caption{Illustration of how the two members of a Cooper pair of electrons (singlet pairing) pick up different contributions $\delta \varphi_{\uparrow}$ and $\delta \varphi_{\downarrow}$ to their phases as a consequence of momentum changes $\delta k_{\uparrow,\downarrow} \sim \pm I/(\hbar v_F) =\pm 1/2\xi_h$ required to conserve energy in response to the magnetic exchange energy shift $\pm I$ in the ferromagnetic weak link. Different paths through the weak link result in different changes to the phase of the pair wave function and, when summed over, tend to suppress the Josephson current.}
\label{Fig8}
\end{figure}

The electrons of a Cooper pair remain coherent while propagating through the normal metal if the extra phase of the pair wave function accumulated along their trajectory can be neglected. Superconducting pairing of electrons with a small total total momentum allows for such a preserved phase coherence over distances of the order of the superconducting coherence length. The situation changes drastically if a ferromagnetic metal is used as the non-superconducting element of a Josephson weak link since the magnetic exchange interaction shifts the energy of electrons with opposite spin projections in different directions, by $\pm I$ say. Since the total energy of both the electrons of a Cooper pair should be conserved after being injected into the ferromagnet the absolute value of their momenta must change, the result being that the total momentum of the injected pair becomes nonzero, $\delta k_{\uparrow,\downarrow} \sim \pm I/(\hbar v_F) = \pm 1/2\xi_h$ (where we have defined the characteristic length $\xi_h=\hbar v_F/2I$ that appears in Fig.~\ref{Fig12}). Consequently, a finite phase is accumulated while the pair is propagating through the magnetic weak link as indicated in Fig.~\ref{Fig8}. 

The phases accumulated by the Cooper pair electrons now depend on their coordinate along the trajectory and are determined by the projection of their velocity on the direction of the supercurrent. To find the total probability amplitude for transferring a Cooper pair from one superconductor to the other across the weak link one needs to sum the phase factors of all possible Cooper pair trajectories. The  destructive interference resulting from such a summation suppresses the Josephson current through a magnetic weak link. \cite{suppression} 

\begin{figure}
\vspace{0.cm} \centerline {\includegraphics[width=9cm]{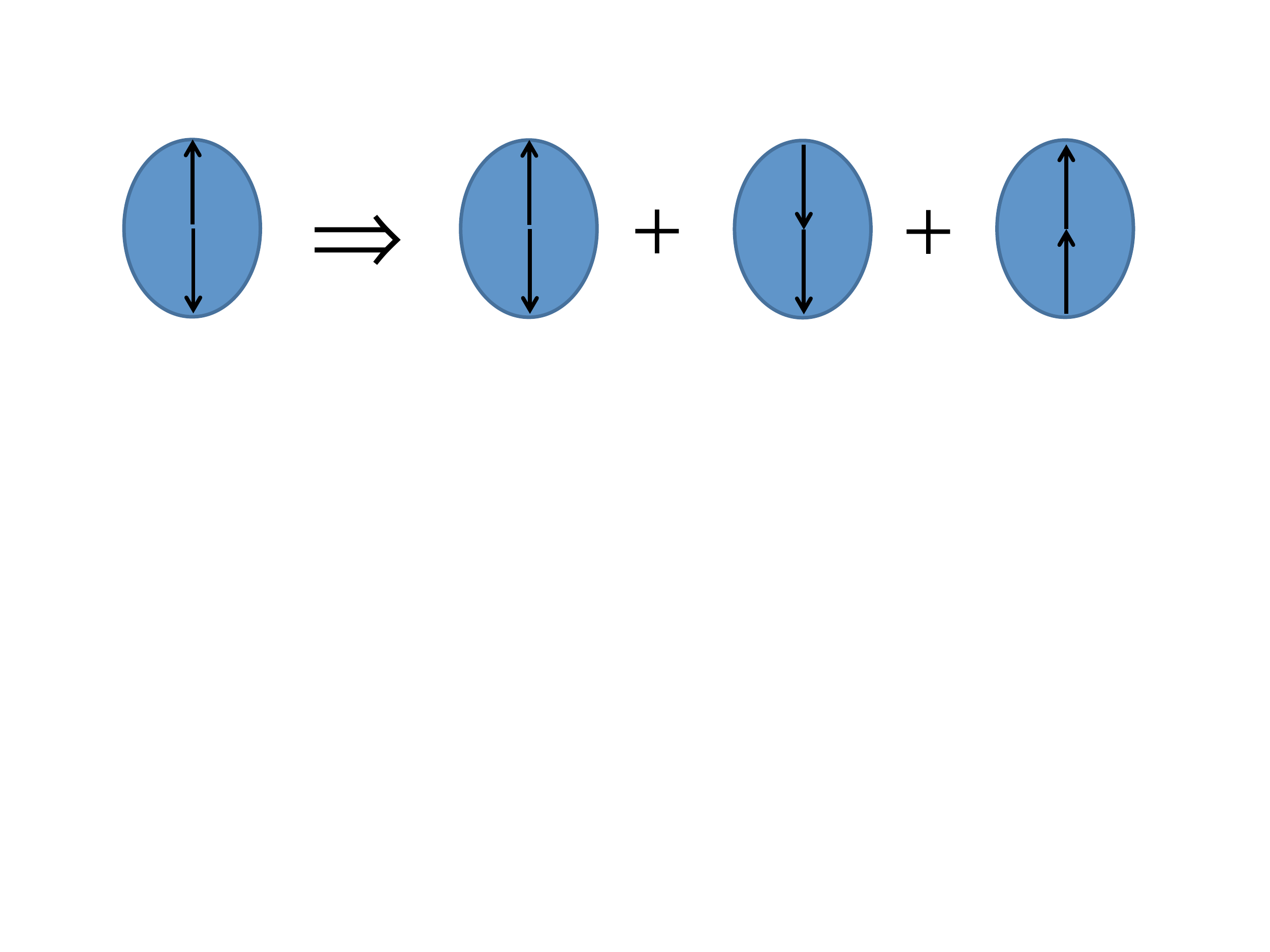}}
\vspace*{-4.0 cm} 
\caption{Schematic representation of how, if the direction of magnetization in the weak link is not constant across the link, quantum fluctuations of the spin lead to a quantum superposition of singlet and triplet paired electrons.}
\label{Fig9}
\end{figure}

If the magnetization direction of the ferromagnetic weak link changes across its length the situation is different since in this case the spin projection 
used to define singlet pairing of the Cooper pair electrons is not a good quantum number and quantum fluctuations of the spin occurs. As a result a quantum superposition 
of singlet and triplet paired electrons occur, as indicated in Fig.~\ref{Fig9}, making it possible for triplet Cooper pairs to contribute to the supercurrent. \cite{Kadig,Volkov}

\begin{figure}
\vspace{0.cm} \centerline {\includegraphics[width=9cm]{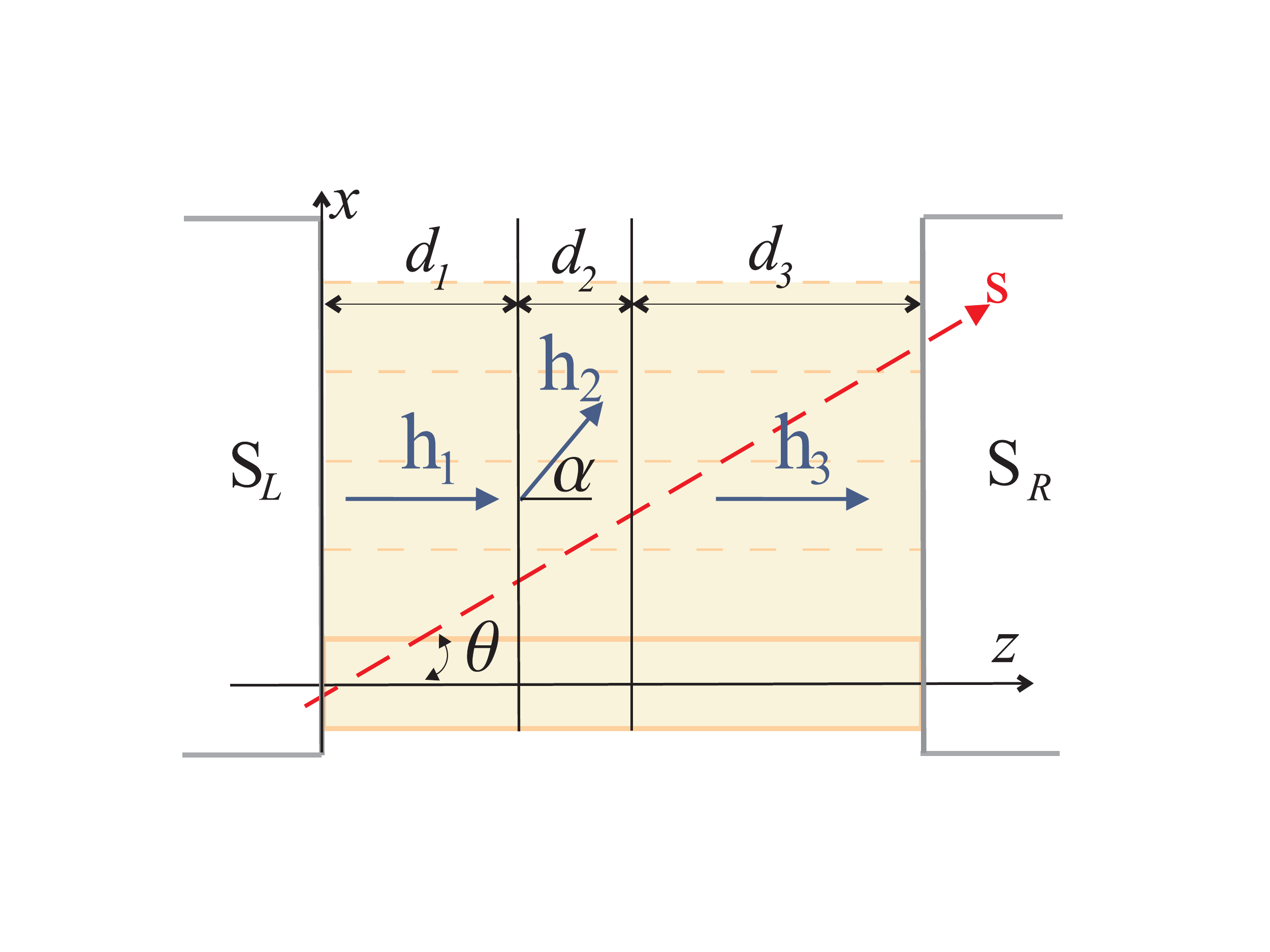}}
\vspace*{-1.0 cm} 
\caption{ 
Model of the S--F--S Josephson junction shown in Fig.~\ref{Fig2}c, here specified to contain three ferromagnetic layers 
(domains) with a stepwise changing profile of the magnetic exchange field ${\bf h}(z)$. The tilt of the exchange field in the middle layer could be due to a nearby magnetic STM tip (spin gate) as in Fig.~\ref{Fig2}c. The tilt angle $\alpha$ is a dimensionless measure of the spin-gate coupling and the dashed red line indicates a linear quasiparticle trajectory.}
\label{Fig10}
\end{figure}

\begin{figure}
\vspace{0.cm} \centerline {\includegraphics[width=9cm]{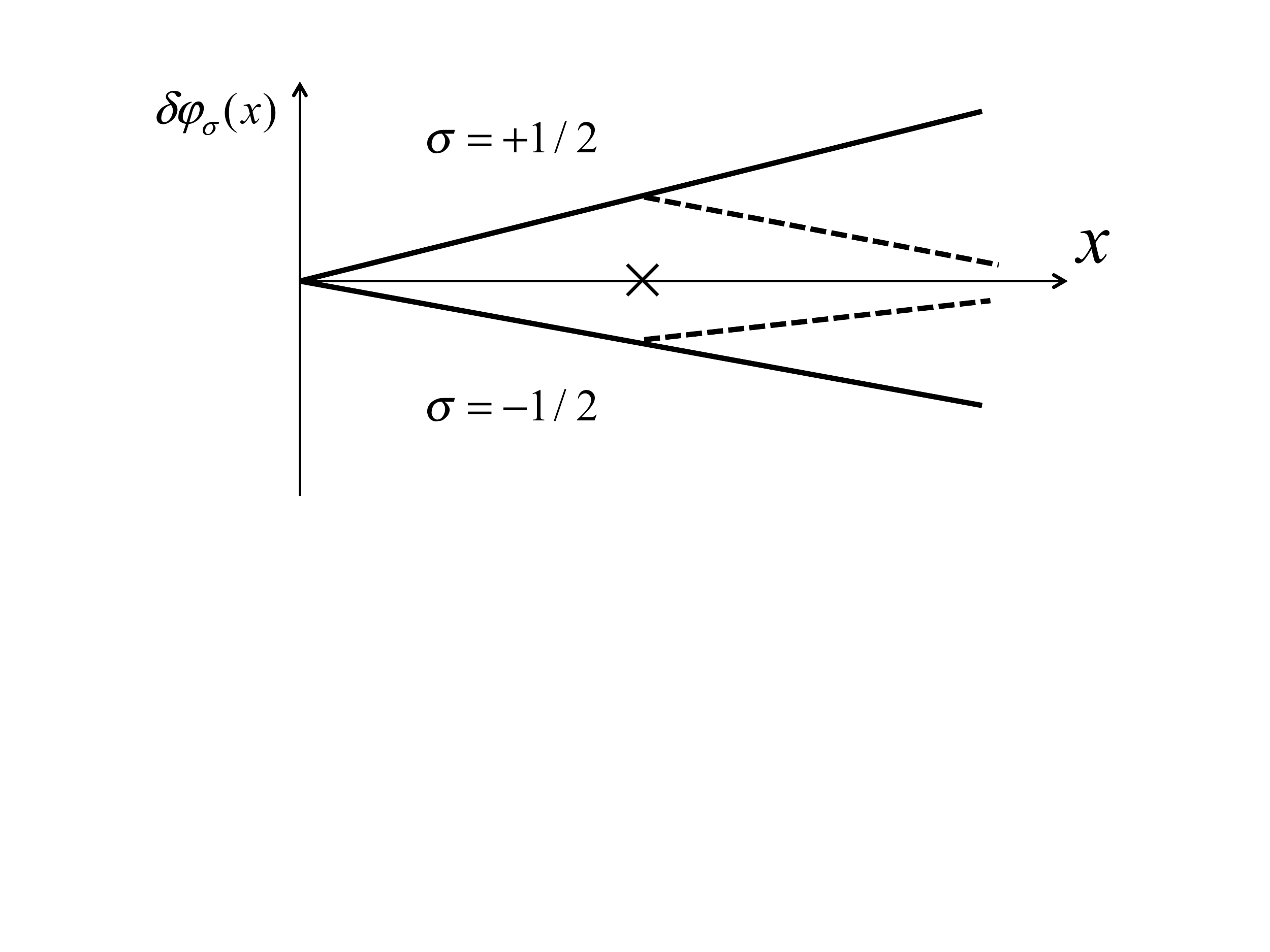}}
\vspace*{-3.0 cm} 
\caption{ 
Illustration of how the phase accumulation $\delta\varphi_{\sigma=\uparrow,\downarrow}$ of the electrons of a Cooper pair as they travel along a trajectory through the model ferromagnetic weak link of Fig.~\ref{Fig10} is reversed after a spin-exchange scattering event.}
\label{Fig11}
\end{figure}

\begin{figure}
\vspace{0.cm} \centerline {\includegraphics[width=9cm]{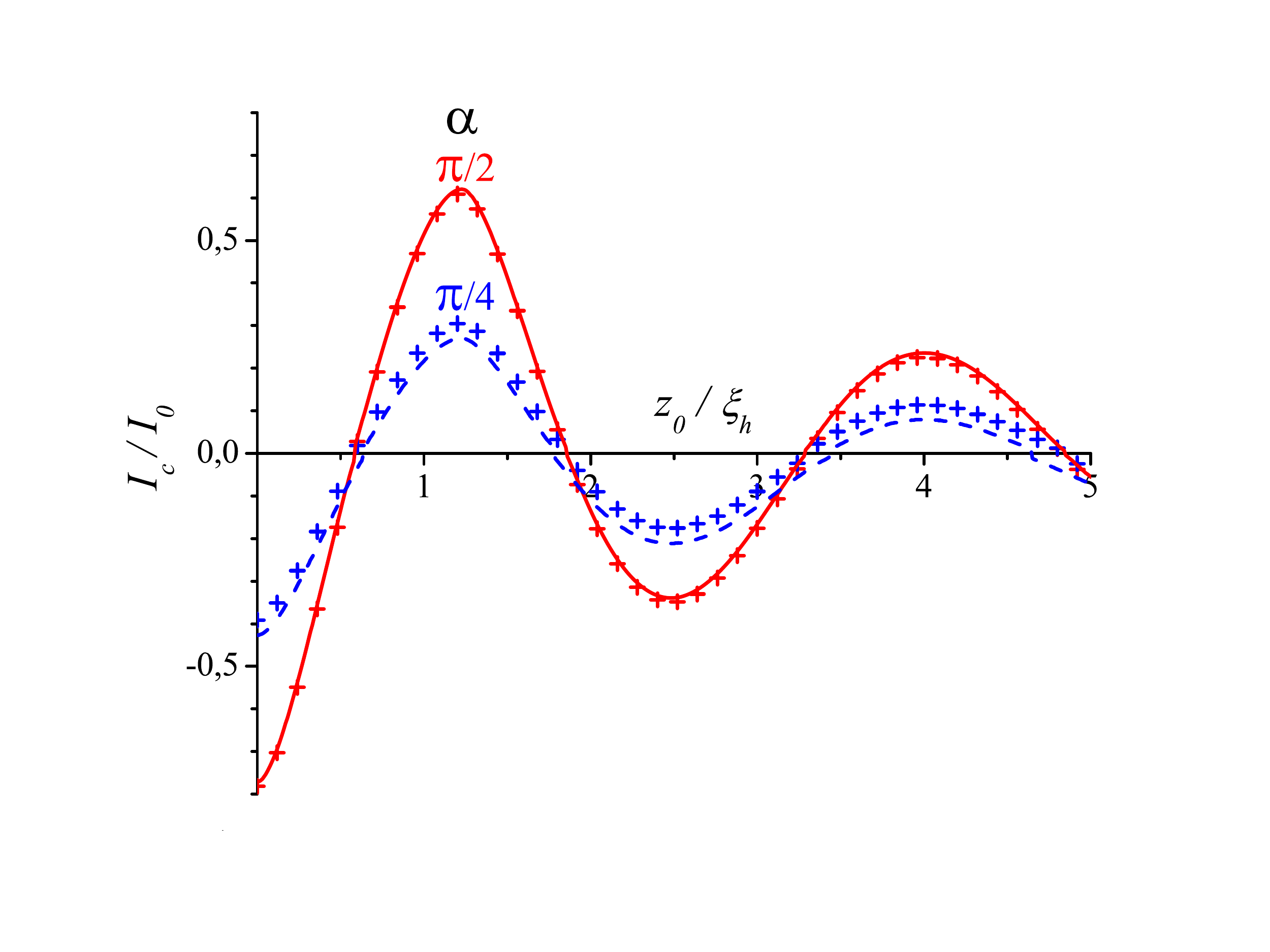}}
\vspace*{-1.0 cm} 
\caption{ 
Dependence of maximal Josephson current $I_c = {\rm max}\{I_1\}$ on the shift of the central domain $z_0$ for different values of the angle $\alpha$ 
We have set $T = 0.9T_c$; $d = 20 \xi_{h}$; $d_2 = 2.5\xi_h$, [$I_0 =(eT_cN/8\hbar)(\Delta/T_c)$].}
\label{Fig12}
\end{figure}

In this Section we consider the ferromagnet-based superconducting weak link sketched in Fig.~\ref{Fig2}c, using the specific model model shown in Fig.~\ref{Fig10}. 
Here the magnetic exchange interaction affects the spin of the Cooper pair electrons as indicated above. In the considered model the effect of spin-gating can be viewed as a local Cooper-pair scattering event occurring in the middle ferromagnetic layer of Fig.~\ref{Fig10}. A possible outcome is an exchange scattering event where the two electrons that form the Cooper pair exchange their spins. In this case  
the phase accumulation of the two paired electrons up to the scattering event, as indicated in Fig.~\ref{Fig8}, will be reversed in the sense that the phase difference between them will diminish after the scattering event as illustrated in Fig.~\ref{Fig11}. This may partially or fully suppress the destructive interference among different Cooper-pair paths through the magnetic weak link and hence increase the effective pair coherence length in the weak link. 



The problem sketched above was considered in Ref.~\onlinecite{Buzdin} by using  a standard parametrization of the anomalous quasiclassical  Green's function $f=f_s+{\bf f}_t\cdot \vec{\sigma}$, which can be viewed as a wave function for the Cooper pairs. Quantum fluctuations of the spin caused by the described spin gating of the device under consideration results in a superposition of singlet ($f_s$) and triplet ${\bf f}_t$ wave functions ($\vec{\sigma}$ is a vector whose components are the Pauli matrices in spin space), which obey the coupled, linearized Eilenberger equations \cite{Eilenberger}
\begin{equation}
\label{EilenbergerEquations}
-i\hbar v_F\partial_s f_s +2{\bf h}\cdot {\bf f}_t =0, \quad -i\hbar v_F\partial_s {\bf f}_t +2{\bf h}f_t =0\,.
\end{equation}
Here $s$ is a coordinate along the Cooper pair trajectory and the boundary conditions at the injection point $s_L$ at the edge of the left superconductor are $f_s(s=s_L)=1$ and ${\bf f}_t(s=s_L)={\bf 0}$.


By solving the set of equations (\ref{EilenbergerEquations}) one can calculate the supercurrent through the system for different electronic trajectories through the weak link (characterized by the angle $\Theta$ shown in Fig.~\ref{Fig10}). One then has to average over $\Theta$ to get the result shown in Fig.~\ref{Fig12}, where the supercurrent is plotted as a function of the position of the spin gate electrode (STM tip) for different (dimensionless) strengths $\alpha$ of the spin-gate coupling (compare Fig.~\ref{Fig10}).

\section{Conclusions}
\label{Conclusions}

The possibility to localize, detect and manipulate single electrons on the nanometer length scale allows for electric charge to be controlled locally down to the level of the fundamental unit of electronic charge. A number of applications have been proposed and realized for such {\em electrostatic gating} of mesoscopic devices. In this review based on recent publications we argue that electronic spin can also be controlled and manipulated locally in a way that provides an additional, spintronic, ``knob" for manipulating nanodevices. Several new functionalities can be achieved by such {\em spin gating} of mesoscopic devices. The magnetic exchange interaction and the Rashba spin-orbit coupling are examples of interactions that can be exploited to provide a spin gating effect. These interactions are both extremely sensitive to geometrical modifications of the device, which makes it possible to provide local probes of electronic spin on the subnanometer length scale. 

A number of suggested new device functionalities based on the spin gating effect on both classical electron transport and quantum coherent electron transport are reviewed in this work. 
Spintromechanics based on the magnetic exchange energy induces a significant spin-polaronic effect in transport through a magnetic nanomechanical SET device and induces a giant spin polarization of the electrical current as described in Section~\ref{Section3A}. Also, a nanomechanical shuttle instability can be induced spintromechanically in such devices as shown in Section~\ref{Section3B}. 

As for spin gating effects in the quantum coherent transport regime, long spin-relaxation times in low-dimensional conductors open up the road to exploring spin-related quantum interference phenomena in mesoscopic devices. The required spin gating of the phase of the electronic wave function can be achieved via the Aharonov-Casher effect, which is to say that if an electron moves in an external electric field an extra spin-dependent phase is accumulated along its trajectory.  In Section~\ref{Section4} we study spin gating effects in a gated nanowire break junction, taking advantage of the fact that the described spin-orbit coupling is greatly enhanced in low dimensional conductors due to the Rashba effect. We demonstrated that in general the junction acts as a ``Rashba spin splitter" by splitting the electronic wave in spin space and that in particular a net spin current can be generated inside the weak link in case a spin imbalance is imposed on the electronic reservoirs coupled by this ``Rashba weak link".
More research is needed to fully develop gate controlled spin interferometry applications, which is likely to involve superconducting Rashba splitters (compare Ref.~\onlinecite{IlyaRashba}) and spin-gated mesoscopic ring configurations.


Finally, in Section~\ref{Section5}, we demonstrate that the phase coherence of a superconducting condensate is sensitive to spin gating through its effect on the Cooper pairs that form the condensate. In particular we showed that weak superconductivity in magnetic Josephson junctions can be stimulated by spin gating of Cooper pairs, which opens up the possibility for spintronic manipulations of superconducting Josephson devices.

With a view towards future work, we note that another possibility to manipulate the electronic spin appears if appropriately designed mesoscopic devices are irradiated by a microwave electromagnetic field. Photon induced electronic spin-flip scattering, e.g., brings a new possibility to affect the nanomechanics of spin-gated devices. This is an interesting future development of spin-gated electronics, which may provide an strong coupling between electromagnetic and mechanical degrees of freedom, allowing for efficient matching of far infrared microwave frequencies with sub-GHz mechanical vibration frequencies. The resonant nature of the photon induced spin flips and spintro-mechanical instability discussed in Section~\ref{Section3B} is a promising feature that may enable a significantly stronger photo-mechanical sub-GHz to sub-THz coupling than is possible with near-equilibrium transducers.

\acknowledgements 
In view of his upcoming retirement from Seoul National University we would like to take this opportunity to thank Professor Yung Woo Park for many years of fruitful scientific collaboration and for hosting us on numerous occasions at various events in Korea. 
Financial support from the Swedish Research Council (VR) is also gratefully acknowledged. 


\begin{thebibliography}{99}
\bibitem{note1}
For the moment we also neglect mechanical damping and intrinsic
spin-flip mechanisms.
\bibitem{instant}
The first non-vanishing term in a perturbation
expansion of the tunneling probability is of order $(\mu H/\Gamma_D)^2$
and corresponds to spin-flip assisted tunneling.
\bibitem{rev2comment}
We note 
the important role of the Coulomb blockade,
which prevents double occupancy of the dot and makes it necessary
for one electron to tunnel to the drain before another can enter
from the source. This means that in between these events the dot
has a deficit of electron charge. Without a Coulomb blockade (as
in Ref.~\onlinecite{Fed2}) there is no restriction on the tunneling sequence
and the dot may have a deficit or a surplus of electrons depending
on the sequence of tunneling to and from the dot. As it turns out
this difference leads to opposite conclusions concerning the
current-induced mechanical stability of the device in Ref.
\onlinecite{Fed2} and this work in the limit of finite electric
force and no exchange interaction.
\bibitem{alternative}
Alternatively, the same result for $U$ can be obtained in the rest frame of the moving magnetic moment by considering its interaction with the magnetic field generated by the current due to the moving charges.
\end{thebibliography}
\end{document}